\documentclass[aps,prl,superscriptaddress,floatfix,twocolumn]{revtex4-2}

\makeatletter
\def\@fnsymbol#1{\ensuremath{\ifcase#1\or *\or \mathsection\or \dagger\or \ddagger\or
    \mathparagraph\or \|\or **\or \dagger\dagger
   \or \ddagger\ddagger \else\@ctrerr\fi}}
\makeatother
\usepackage{amsmath}
\usepackage{amssymb}
\usepackage{amsfonts,mathtools}
\usepackage{graphicx}
\usepackage{xcolor}
\usepackage[colorlinks,linkcolor=blue,urlcolor=blue,citecolor=blue]{hyperref}
\usepackage{float}

\usepackage[all]{hypcap}
\newcommand{\beq}{\begin{equation}}
\newcommand{\eeq}{\end{equation}}
\usepackage{comment}

\newcommand{\D}{\mathcal{D}}
\newcommand{\SFF}{\mathrm{SFF}}

\begin{document}

\title{Airy distribution of the quenched spectral form factor\\ in non-interacting fermions}

\author{Oscar Martin}
\affiliation{IPhT, CNRS, CEA, Universit\'e Paris Saclay, 91191 Gif-sur-Yvette, France}
\affiliation{École Normale Sup\'erieure Paris-Saclay, Universit\'e Paris-Saclay, 91190 Gif-sur-Yvette, France}

\author{Satya N. Majumdar}
\affiliation{Universit\'{e} Paris Saclay, CNRS, LPTMS, 91405, Orsay, France}

\author{Laura Foini}
\email{laura.foini@ipht.fr}
\affiliation{IPhT, CNRS, CEA, Universit\'e Paris Saclay, 91191 Gif-sur-Yvette, France}

\date{\today}

\begin{abstract}
For chaotic many-body systems as well as the random energy model, the logarithm of the (scaled)
spectral form factor (SFF) exhibits two robust properties at late times: (i) self-averaging 
and (ii) a reflected-Gumbel distribution. Here, we demonstrate that both properties break 
down for $N$ non-interacting fermions on a lattice of size $L$ with an incommensurate 
single-particle spectrum. We show that the logarithm of the SFF, upon rescaling by 
$1/\sqrt{L}$, follows the Airy distribution, in stark contrast to the reflected-Gumbel law. 
Moreover, it is no longer self-averaging, as its mean and standard deviation are of the same 
order. While deviations from reflected-Gumbel statistics are generically expected for 
non-interacting systems, we find that an \emph{interacting integrable} model exhibits reflected-Gumbel 
statistics at late times, mirroring the chaotic case and distinguishing it from 
non-interacting fermions.
\end{abstract}

\maketitle

\emph{Introduction.} The spectral form factor (SFF) is a central diagnostic of the correlations of quantum spectra, from single-particle chaotic systems to many-body Hamiltonians and holographic models \cite{Leviandier1986,Berry1985,SieberRichter2001,Haake2018,Cotler2017,PhysRevLett.121.060601,PhysRevE.55.4067,PhysRevResearch.3.L012019}. 
In the canonical ensemble
it is defined as the squared modulus of the analytically continued partition function $\SFF_N(t) = |Z_N(\beta+it)|^2$ with $Z_N(\beta+it) = \mathrm{Tr}_N\, e^{-(\beta+it) H}$, the trace running over the sector with $N$ particles, resolving the spectrum in time.

In chaotic systems, spectral correlations combine strong local level repulsion with the long-range universal correlations of random-matrix theory. These correlations produce the celebrated slope-dip-ramp-plateau structure of the \emph{average} SFF, while for uncorrelated (Poissonian) levels the ramp is absent.

Beyond its average, the SFF is well known not to be self-averaging \cite{Prange1997}: at long times its fluctuations remain of the order of its mean, whatever the system size. Its logarithm - the \emph{quenched} SFF, in the language of disordered systems - is instead self-averaging, with bounded and universal fluctuations \cite{Charamis2026}. 
The reason is that $Z_N(\beta+it)$ becomes a complex Gaussian variable at long times \cite{Cipolloni2023,Altland2025,Charamis2026}: despite the spectral correlations, it behaves as if the central limit theorem applied to its exponentially many terms. Its squared modulus, properly normalized, is then exponentially distributed, and its logarithm follows a reflected Gumbel distribution, whose exponential left tail encodes the near misses of the Fisher zeros in the complex temperature plane \cite{Bunin2024,Charamis2026,Derrida1991}.

At fixed spectrum and $\beta=0$, a simple sufficient condition makes this transparent. If the $\D$ many-body energies $E_n$ are rationally independent~\footnote{Real numbers $x_1,\dots,x_K$ are \emph{rationally independent} if no nontrivial
integer combination of them vanishes: $\sum_{k=1}^K m_k x_k=0$ with $m_k\in\mathbb{Z}$
implies $m_1=\dots=m_K=0$.}, the Kronecker-Weyl theorem \cite{Kac1959,KuipersNiederreiter} implies that sampling $t$ uniformly at long times is equivalent to drawing the phases $E_n t \bmod 2\pi$ independently and uniformly on $[0,2\pi)$. The time average thus erases the spectral correlations, whatever they are, and $Z_N(it)=\sum_{n=1}^{\D} e^{-iE_n t}$ becomes a sum of $\D$ independent terms, to which the central limit theorem applies.

This reflected-Gumbel statistics is remarkably robust. It holds in chaotic systems as well as in the random energy model, whose levels are independent \cite{Charamis2026}, and we show below that it also holds for \emph{interacting} integrable models: on the plateau, the long-time distribution of the SFF does not tell them apart from chaotic ones. What this scenario requires is that the $\D$ many-body levels constitute as many independent degrees of freedom, as guaranteed by their rational independence.

Non-interacting particles violate precisely this assumption. In this work we show that, when their single-particle spectrum is incommensurate, they escape the reflected-Gumbel law altogether, and fall into an entirely different universality class, with no adjustable parameter and independent of the microscopic details of the model.

For $N$ noninteracting fermions in $L$ single-particle levels, the $\D=\binom{L}{N}$ many-body levels are not independent: each of them is the plain sum of $N$ out of only $L$ single-particle energies, so that the $E_n$ obey a host of integer relations and the Kronecker-Weyl theorem applies only to the $L$ single-particle phases $e^{-i\varepsilon_j t}$, provided they are rationally independent, not to the $\D$ many-body ones.
 At fixed density $\rho=N/L$ and at long times, we find that this redundancy changes both the form and the scale of the fluctuations of the quenched SFF:
\beq
\frac{1}{2\pi \sqrt{L}} \ln \SFF_N \to \, M , \quad M \sim f_{\rm Airy},
\label{eq:finalresult}
\eeq
where $f_{\rm Airy}$ is the Airy distribution [Eq.~\eqref{eq:airy} below], with no adjustable parameter. The fluctuations of $\ln\SFF_N$ are therefore of order $\sqrt{L}$, instead of the $O(1)$ Gumbel fluctuations of the chaotic case: the logarithm of the SFF is no longer self-averaging. 
The quenched average $\langle\ln\SFF_N\rangle\simeq 2\pi\sqrt{\pi/8}\,\sqrt{L}$ is moreover parametrically smaller than the annealed one, $\ln\langle\SFF_N\rangle=\ln\D\propto L$ \footnote{By Stirling's formula, at fixed density $\rho=N/L$,
$\ln\D=\ln\binom{L}{\rho L}=L\,s(\rho)+O(\ln L)$,
with $s(\rho)=-\rho\ln\rho-(1-\rho)\ln(1-\rho)$ the entropy per site; at half filling $s(1/2)=\ln2$.}: quenched and annealed averages differ at leading order, whereas for a reflected Gumbel distribution they differ only by a constant, $\langle\ln\SFF_N\rangle=\ln\D-\gamma$.

We obtain this result by introducing an effective landscape defined on the phase
$\theta$ conjugate to the particle number. Although deterministic at fixed disorder,
at long times this landscape behaves as a periodic Brownian path, with Brownian
increments and an area that vanishes exactly for every realization. In the
grand-canonical ensemble the SFF probes this landscape at a single point and is
log-normal~\cite{Venuti2025}; fixing $N$ instead selects its maximum, so that
typical-value statistics are traded for extreme-value statistics. The maximum of a
zero-area periodic Brownian path is the maximal relative height of a fluctuating
periodic interface, Airy distributed according to Majumdar and
Comtet~\cite{majumdar2004,majumdar2005}.

The Airy distribution is remarkably ubiquitous. Originally obtained for the area under a Brownian excursion
\cite{Louchard1984,Takacs1991}, the Airy distribution also governs the maximum of
a Brownian bridge measured from its mean \cite{Darling1983}, and appears in
combinatorics and computer science, from the cost of linear-probing hashing to the enumeration of connected graphs \cite{FlajoletLouchard2001,Janson2007}, and in statistical physics as the law of the maximal height of interfaces in the stationary state \cite{majumdar2004,majumdar2005,SchehrMajumdar2006}. Here it emerges from a \emph{deterministic} signal: at fixed disorder, $\SFF_N(t)$ is a fixed function of time, and randomness is generated solely by the incommensurability of the single-particle energies, in the spirit of Kac's statistical independence \cite{Kac1959}. 

This places our result within a broader family of pseudo-random problems, such as the value distribution of the Riemann zeta function, whose statistics under time sampling are controlled by the phases $p^{-it}$, effectively independent owing to the rational independence of the logarithms of the primes \cite{Soundararajan2022,BerryKeating1999,LeClairMussardo2019}.

The quenched SFF of free particles provides a physical instance of such a pseudo-random process in which, owing to the projection onto fixed $N$, the relevant statistics are those of an extreme value rather than of a typical one.

Two ingredients enter, and only two: the many-body spectrum must be \emph{additive}, built out of $L$ single-particle energies, and those single particle energies must be \emph{incommensurate}, i.e.\ rationally independent. Neither dimensionality, nor the localized or delocalized character of the eigenstates plays any role. The result covers disordered and quasiperiodic free-particle systems alike.

\emph{Setting.} We consider $N$ noninteracting fermions. Let $L$ denote the number of single-particle levels, with energies $\varepsilon_i$, $i=1,2,\dots,L$. We place the $N$ fermions in these $L$ levels according to the Pauli exclusion principle, namely no two fermions can occupy the same level. If $n_i=0,1$ denotes the occupancy of level $i$, any many-body energy level $E$ is labelled by a configuration $\{n_i\}$ satisfying the two constraints $E=\sum_{i=1}^L n_i\varepsilon_i$ and $N=\sum_{i=1}^L n_i$. There are a total of $\D=\binom{L}{N}$ many-body levels, and consequently the canonical partition function is given by
\beq
Z_N(\beta+it) = \sum_{\{n_i=0,1\}} e^{-(\beta+it)\sum_{i} n_i\varepsilon_i}\,\delta_{\sum_{i} n_i,N}\,.
\label{eq:ZN}
\eeq
We assume only the two properties stated above: the many-body spectrum is additive, and the $\varepsilon_j$ are rationally independent - which holds for any continuous distribution and, more generally, for generic quasiperiodic potentials. Everything below depends on the microscopic model through these two properties alone. The particle number is conserved and we work at fixed density $\rho = N/L$. The $\D$ levels have Poissonian spacings but, built from only $L$ numbers, are strongly correlated on larger spectral scales - a correlation invisible to the gap ratio, and precisely what the long-time distribution of the SFF sees.

Multiplying both sides of Eq.~\eqref{eq:ZN} by $z^N$ and summing over $N=0,1,\dots,L$ lifts the Kronecker delta, and the sum over the independent $n_i$ factorizes, giving the generating function
\beq
\begin{split}
P(z) &= \sum_{N=0}^{L} Z_N(\beta+it)\, z^N = \sum_{\{n_i=0,1\}} \prod_{j=1}^{L} \left(z\,e^{-(\beta+it)\varepsilon_j}\right)^{n_j}\\
&= \prod_{j=1}^{L} \left(1 + z\, e^{-(\beta+it)\varepsilon_j}\right),
\end{split}
\label{eq:grandcanonical}
\eeq
a polynomial in $z$, so that $Z_N$ is the elementary symmetric polynomial of the $\{e^{-(\beta+it)\varepsilon_j}\}$. From now on we set $\beta=0$ (see the conclusions). By Cauchy's formula on the unit circle $z=e^{i\theta}$,
\beq
\begin{split}
Z_N(it) &= \frac{1}{2\pi} \int_{0}^{2\pi}\! P(e^{i\theta})\, e^{-i N \theta}\, d\theta\\
&= \frac{1}{2\pi} \int_{0}^{2\pi}\! \prod_{j=1}^{L} \left(1 + e^{i\theta} e^{-i\varepsilon_j t}\right) e^{-i N \theta}\, d\theta .
\end{split}
\label{eq:cauchy}
\eeq
This exact projection onto the fixed-number sector is our starting point: the whole time dependence enters through the $L$ phases $e^{-i\varepsilon_j t}$ on the unit circle, and $\theta$ is the collective parameter announced above.

\begin{figure}[h!]
\includegraphics[width=\columnwidth]{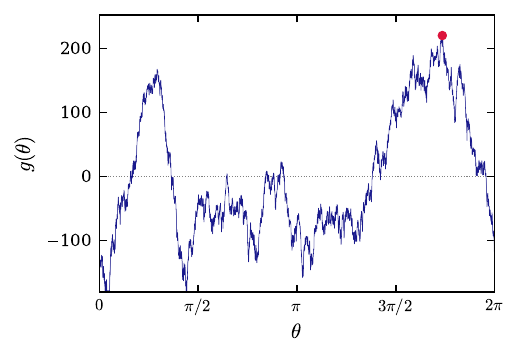}
\caption{The landscape $g(\theta)$, Eq.~\eqref{eq:gdef}, for a single disorder realization of Eq.~\eqref{eq:hamiltonian} with $L=10^4$, $h_i$ uniform in $[-1,1]$, at $t=10^6$. The smooth envelope is a Brownian-like path of amplitude $\sim\sqrt L$ oscillating about zero - the exact zero-area constraint \eqref{eq:zeroarea} - decorated by thin integrable logarithmic spikes at the zeros $\theta_k$ of $P$ on the unit circle. The red dot marks the global maximum, which controls the quenched SFF through Eq.~\eqref{eq:lnZmax}.}
\label{fig:gtheta}
\end{figure}

\emph{Long times.} The phases $\phi_j(t)=-\varepsilon_j t \bmod 2\pi$ wind at incommensurate rates. 
If the $\varepsilon_j$ are rationally independent, as in a typical disordered case, the Kronecker-Weyl theorem \cite{Kac1959,KuipersNiederreiter} makes $(\phi_1,\dots,\phi_L)$ equidistributed on the torus $\mathbb{T}^L$ (see Supplemental Material~\cite{SM}). This is the same theorem invoked in the introduction, now applied to the $L$ single-particle energies rather than to the $\D$ many-body ones. Sampling $t$ uniformly on $[0,T]$, $T\to\infty$, is then equivalent to
\beq
e^{-i\varepsilon_j t} \;\longrightarrow\; a_j = e^{i\phi_j}\,, \qquad \phi_j \sim \mathcal{U}([0,2\pi))\;\; \text{i.i.d.}
\label{eq:iidphases}
\eeq
The same mechanism underlies the plateau of the averaged SFF; here we keep the full distribution induced by the random phases, not just its first two moments. Writing the integrand of \eqref{eq:cauchy} as $e^{g(\theta)+i\Psi(\theta)}$, the modulus is controlled by
\beq
g(\theta) = \ln\big|P(e^{i\theta})\big| = \sum_{j=1}^{L} \ln\big|1 + a_j e^{i\theta}\big| ,
\label{eq:gdef}
\eeq
the random landscape of Fig.~\ref{fig:gtheta}. Unitarity makes the phase $\Psi$ piecewise linear with constant slope $L(\tfrac12-\rho)$, jumping by $\pi$ at each of the $L$ points $\theta_k=\pi-\phi_k$ where $P$ vanishes; these zeros all lie on the unit circle. The modulus of the integrand is $e^{g(\theta)}$, so that
$|Z_N|\le e^{\max_\theta g}$. This bound is saturated at leading order.
The $L$ zeros of $P$ on the unit circle, spaced by $\sim 1/L$, make each
peak of $e^{g}$ so narrow that the phase of the integrand turns by only
$O(1)$ across it and cannot cancel it; and since the peak heights differ
by factors up to $e^{O(\sqrt{L})}$, the highest one dominates the
contributions of all others. One therefore finds:
\beq
\ln|Z_N| = \max_{\theta} g(\theta) + O(\ln L),
\label{eq:lnZmax}
\eeq
where the estimated logarithmic correction  originates from the $1/L$ width of the dominant arc between two zeros~\cite{SM}.

\emph{A zero-area Brownian bridge.} The problem is thus one of extreme
values for $g$. We denote by $\langle\bullet\rangle$ the average over the
i.i.d.\ uniform phases $\phi_j$ of Eq.~(\ref{eq:iidphases}), which by Weyl's theorem is
equivalent to the long-time average at fixed disorder. Each term of
Eq.~(\ref{eq:gdef}) has zero mean, since shifting $\theta$ only shifts a uniform
phase and $\int_0^{2\pi}\ln|1+e^{ix}|\,dx=0$; hence
$\langle g(\theta)\rangle=0$ for every $\theta$. The fluctuations are
characterized by the two-point function
$C(\tau)=\langle g(\theta)\,g(\theta+\tau)\rangle$, which by the same
argument does not depend on $\theta$: the process is stationary. Since
the $\phi_j$ are independent, only the diagonal terms $j=k$ survive, and
inserting the Fourier series
$\ln|1+e^{ix}|=\sum_{n\ge1}\frac{(-1)^{n-1}}{n}\cos(nx)$, orthogonality
of the cosines gives~\cite{SM}:
\begin{equation}\label{eq:covariance}
C(\tau)=\frac{L}{2}\sum_{n\ge1}\frac{\cos(n\tau)}{n^2}
=L\left(\frac{\pi^2}{12}-\frac{\pi}{4}|\tau|+\frac{\tau^2}{8}\right),
\end{equation}
for $|\tau|\le2\pi$, i.e.\ $\pi^2L/2$ times the periodic Bernoulli
polynomial $B_2(\tau/2\pi)$. In particular
$\langle g^2\rangle=C(0)=\pi^2L/12$: the landscape $g(\theta)$ fluctuates on the
scale $\sqrt{L}$.
Three properties identify the effective process. (i) By the multivariate central limit theorem all finite-dimensional marginals of $g/\sqrt L$ converge to Gaussians with covariance \eqref{eq:covariance}; $g$ itself is not Gaussian at finite $L$, because of the logarithmic divergences at the $\theta_k$. (ii) Increments are locally Brownian, $\mathrm{Var}[g(\theta+\delta)-g(\theta)] \simeq \pi L |\delta|/2$ for $\delta\ll1$, with a weak uniform anticorrelation between disjoint increments. (iii) That anticorrelation implements an exact constraint in fact one has:
\beq
\frac{1}{2\pi} \int_0^{2\pi} g(\theta)\, d\theta = 0
\label{eq:zeroarea}
\eeq
\emph{identically, for every realization} of the phases. The landscape is a periodic Brownian-like path of exactly vanishing area.

The identification then follows from the covariance alone. With $u=\theta/2\pi$, the circular Brownian path of unit period and zero area, $\mathcal{B}_0(u)=B(u)-\int_0^1 B$, with $B$ a standard bridge, is stationary with $\mathrm{Cov}[\mathcal{B}_0(u),\mathcal{B}_0(u+s)]=B_2(s)/2$, which for $s=\tau/2\pi$ is exactly $\pi^2L$ times smaller than \eqref{eq:covariance}. Two Gaussian processes with the same covariance being equal in law, $g(2\pi u)\to\pi\sqrt{L}\,\mathcal{B}_0(u)$; the convergence holds for the whole process and hence it is expected for its maximum, the integrable downward divergences leaving it unaffected \cite{TucciWhiting2014}. The limit is the stationary height profile, measured from its spatial average, of a one-dimensional Edwards-Wilkinson or KPZ interface of length $\ell$, $\tilde h(x)=\sqrt{\ell}\,\mathcal{B}_0(x/\ell)$, with covariance $\tfrac{\ell}{2}B_2(x/\ell)$ at separation $x$; under the dictionary $\ell=\pi^2L$, $2\pi x/\ell=\tau$ this is \eqref{eq:covariance} term by term. 

The maximum of $g/(\pi\sqrt{L})$ is therefore the Majumdar-Comtet observable, the maximal relative height of such an interface in units of $\sqrt{\ell}$, which is Airy distributed \cite{majumdar2004,majumdar2005}:
\beq
f_{\rm Airy}(y) = \frac{2\sqrt{6}}{y^{10/3}} \sum_{n=1}^{\infty} e^{-b_n/y^2}\, b_n^{2/3}\, U\!\left(-\tfrac{5}{6}, \tfrac{4}{3}, \tfrac{b_n}{y^2}\right),
\label{eq:airy}
\eeq
with $b_n=2\alpha_n^3/27$, $\alpha_n$ the moduli of the zeros of the Airy function and $U$ the confluent hypergeometric function; its mean is $\sqrt{\pi/8}$ and its standard deviation $\sigma_{\rm Airy}=\sqrt{5/12-\pi/8}$. With \eqref{eq:lnZmax} and $\SFF_N=|Z_N|^2$, we obtain the result announced in Eq.~\eqref{eq:finalresult}.

The contrast with the chaotic and many-body Poissonian (random energy model) cases \cite{Charamis2026} is striking. There the quenched average follows the annealed one, $\langle \ln \SFF_N\rangle \simeq \ln\D \propto L$ on the plateau, with $O(1)$ Gumbel fluctuations of variance $\pi^2/6$. Here the annealed average is still $\ln\langle\SFF_N\rangle = \ln\D \propto L$ - the plateau of $\langle\SFF_N\rangle$ is exactly $\binom{L}{N}$, only diagonal pairings surviving the phase average - whereas the quenched average grows only as $2\pi\sqrt{\pi/8}\sqrt{L}$, parametrically smaller, with fluctuations of the same order $\sqrt L$. Quenched and annealed averages therefore differ at leading order, as anticipated in the introduction. By contrast, summing over all particle numbers gives $\sum_N Z_N=P(1)$, so that the grand-canonical form factor obeys $\ln\SFF_{\rm gc}=2g(0)$: a typical value  of the landscape rather than its maximum.

\emph{Numerics.} We now specialize to a concrete realization of the above class, the one-dimensional Anderson model, written as an XX chain in a random longitudinal field,
\beq
H = \sum_{i=1}^{L} \left( S_{i}^{x}S_{i+1}^{x} + S_{i}^{y}S_{i+1}^{y} \right) + \sum_{i=1}^L h_{i}\, S_{i}^{z}\,,
\label{eq:hamiltonian}
\eeq
with $h_i$ i.i.d.\ uniform on $[-W,W]$ and $L$ being the number of sites. A Jordan-Wigner transformation maps $H$ onto free spinless fermions in a random potential, with Anderson-localized eigenstates \cite{Anderson1958}. The result is independent of $W$, which enters only to guarantee incommensurability. We use this model because it makes incommensurability manifest and is tractable at the large sizes the $\sqrt L$ scaling demands. Note that no disorder average is involved: at fixed disorder the spectrum is frozen and $\SFF_N(t)$ is a deterministic function, the randomness coming entirely from sampling $t$ at long times.

Figure~\ref{fig:gtheta} summarizes the mechanism. The envelope wanders on the $\sqrt L$ scale predicted by \eqref{eq:covariance} and oscillates about zero, positive and negative excursions compensating exactly as imposed by the sum rule \eqref{eq:zeroarea}, which the data verify to machine precision. The logarithmic singularities form a dense fringe of thin downward spikes of typical spacing $2\pi/L$ that cannot interfere with the maximum. That maximum sits on a smooth arc, at the top of the largest excursion; $g$ at a generic angle would give only the much smaller typical value.

\begin{figure}[h!]
\includegraphics[width=\columnwidth]{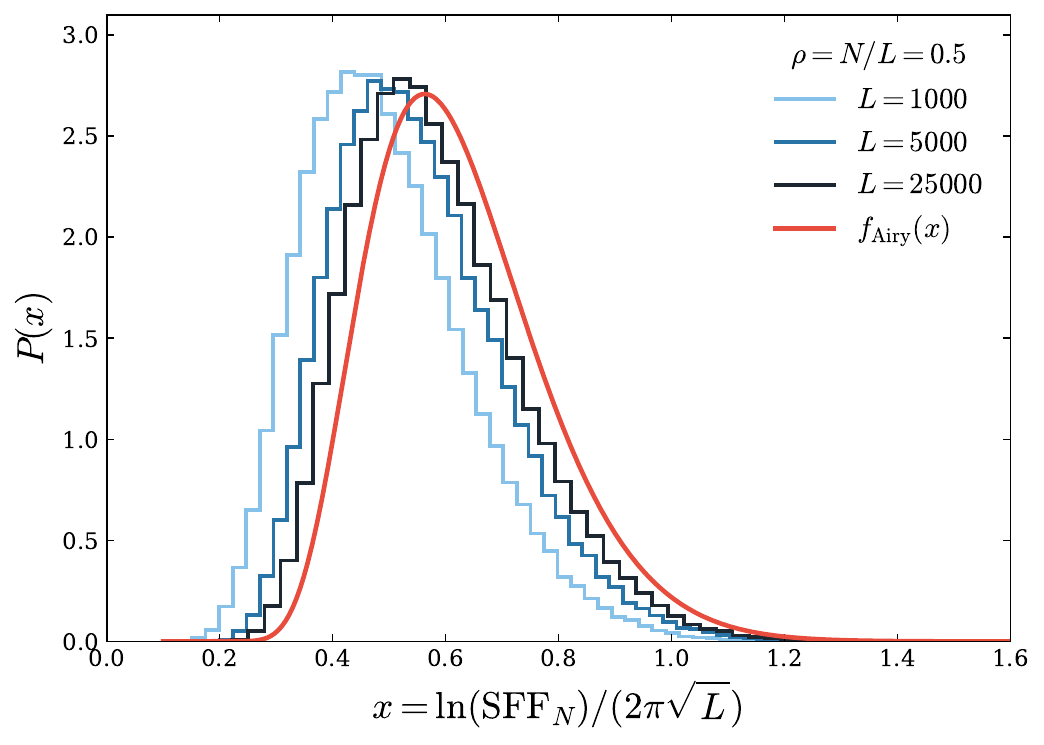}
\caption{Empirical distribution of $x = \ln \SFF_N/(2\pi\sqrt{L})$ for $L = 1000$, $5000$ and $25000$ ($10^5$ samples of $t\in[10^3,10^8]$ each, $\rho=1/2$, $W=1$). The red curve is $f_{\rm Airy}$, Eq.~\eqref{eq:airy}, with no adjustable parameter: the normalization $2\pi\sqrt L$ is fixed microscopically by the covariance \eqref{eq:covariance}. The shape and width converge rapidly, the mean slowly.}
\label{fig:airycollapse}
\end{figure}

The histograms of Fig.~\ref{fig:airycollapse},
as a function of the rescaled variable $x=\ln\SFF_N/(2\pi\sqrt{L})$,
show the characteristic Airy shape already at the smallest sizes: strongly asymmetric, with the essential-singularity suppression $\sim e^{-b_1/x^2}$ of low maxima and a right tail decaying faster than exponentially. We observe a slow convergence of the mean which is the signature of the subleading $O(\ln L)$ term in \eqref{eq:lnZmax}. 

We note that the Airy law is supported on $x\ge 0$: by the zero-area constraint
\eqref{eq:zeroarea}, $\max_\theta g(\theta)\ge 0$ for every realization.
The zeros of $Z_N$ in the complex-temperature plane, whose near misses
produce the exponential left tail of the Gumbel law in the chaotic case,
should only affect the $O(1)$ correction to
Eq.~\eqref{eq:finalresult}.

\emph{Interactions.} What produces the Airy law is the free-particle structure of the spectrum, and not the integrability of the model. To make this explicit we turn to the XXZ chain:
\begin{equation}
H=\sum_{i=1}^{L} ( S^x_iS^x_{i+1}+S^y_iS^y_{i+1}+\Delta S^z_iS^z_{i+1}) \, 
\end{equation}
which for $\Delta\neq0$ describes \emph{interacting} fermions. It is Bethe-integrable and its level spacings are Poissonian, as in the Anderson model; but its eigenenergies are sums of quasiparticle energies whose rapidities are dressed by the interaction in a way that depends on the whole state, so the many-body levels are no longer rigidly built from a fixed set of $L$ numbers. 
There is no analogue of \eqref{eq:grandcanonical}, and no integer relations are expected among the many-body energies beyond those imposed by symmetries.  The central limit theorem is then restored, and the variable
\beq
Y=\ln\frac{\SFF_N(t)}{\langle\SFF_N(t)\rangle}, \label{eq:Ydef}
\eeq
sampled at long times, where the SFF fluctuates around its plateau value,
should follow the reflected Gumbel law $P(y)=e^{y-e^{y}}$, exactly as in the chaotic case.
The integrable or chaotic nature of the model is encoded in the time dependence
of $\langle\SFF_N(t)\rangle$, which may or may not display a ramp at intermediate
times, but not in the long-time distribution of its logarithm.

\begin{figure}[h!]
\includegraphics[width=\columnwidth]{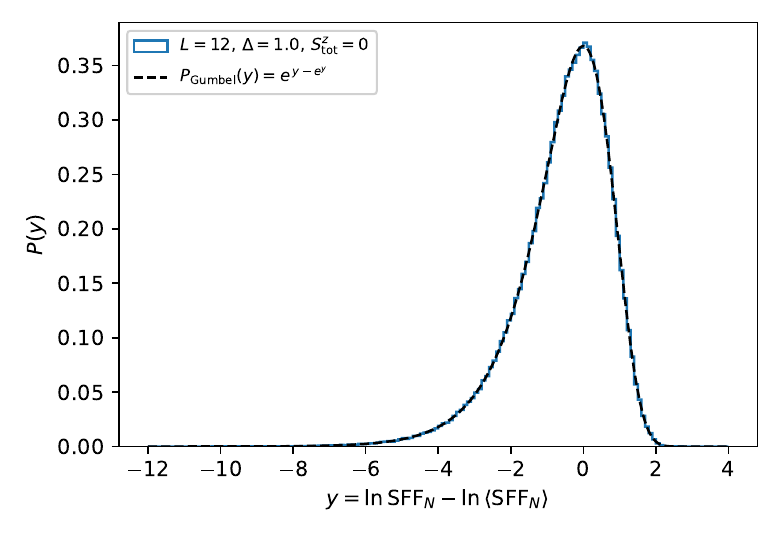}
\caption{Distribution of $Y=\ln\SFF_N -\ln\langle\SFF_N\rangle$ for the clean XXZ chain with $L=12$, $\Delta=1$, at half filling, $\D=924$. The SFF is sampled at $4\times10^{5}$ random times in $t\in[10^{5},2\times10^{5}]$ and $\langle\SFF_N\rangle$ is the empirical time average. The dashed line is $P(y)=e^{y-e^{y}}$, with no adjustable parameter: for this interacting integrable model the distribution of the quenched SFF behaves exactly as in the chaotic case.}
\label{fig:xxzgumbel}
\end{figure}

This is what we find, and remarkably no disorder is needed - the time average suffices. Figure~\ref{fig:xxzgumbel} shows $Y$ for the \emph{clean} chain at $\Delta=1$ and half filling, sampled at late times in the plateau of a single Hamiltonian: it converges to a reflected Gumbel with no adjustable parameter, mean $-\gamma$ and variance $\pi^2/6$, in sharp contrast with the Airy law of width $\sqrt L$ of the free case.

\emph{Conclusions.} For free particles with an incommensurate spectrum at fixed density, the quenched SFF has anomalously large, non-Gumbel fluctuations: $\ln\SFF_N$ follows the Airy law of the maximum of a periodic zero-area Brownian path, on the scale $\sqrt L$. The mechanism is transparent: the analytically continued partition function is a random polynomial in only $L$ phases, and its logarithm is set by the maximum of the correlated landscape $g$, whose Brownian scaling and exact zero-area constraint produce the Majumdar-Comtet universality. The distribution of the quenched $\SFF_N$ is therefore a sharp long-time probe of the free-particle structure of the many-body spectrum - exponentially many levels being rigid combinations of $L$ energies - invisible to short-range level statistics, Poissonian in both the free and the interacting integrable case.

Several directions follow. 
At finite temperature the zeros of $P$ move off the unit circle and its Fourier modes are damped. One therefore expects a crossover from the Airy
law to other non-universal extreme-value statistics.  
Since only additivity and incommensurability enter, the Airy law should survive in higher dimension and for delocalized disordered free systems; whether the localized and delocalized phases of the Anderson model in $d\ge3$ can be told apart at the level of finite size corrections is open. Quasiperiodic systems, where incommensurability is built in rather than supplied by disorder, would test the mechanism in a fully deterministic setting. Finally, \emph{bosons} put the mechanism to a sharp test: the generating function becomes $\prod_j(1-z e^{-i\varepsilon_jt})^{-1}$, whose poles again lie on the unit circle, so the contour must be taken at $r<1$; 
whether the Airy distribution applies is left open.

\emph{Acknowledgements.}
This work was supported by the French government through the France 2030 program (PhOM - Graduate School of Physics), under reference ANR-11-IDEX-0003 (Project Mascotte, L. Foini).
S. N. M. acknowledges support from the grant ANR-23-CE30-0020-01 EDIPS.

\bibliography{biblio}

\newpage

\onecolumngrid
\vspace{1.2em}
\begin{center}
\rule{0.42\textwidth}{0.4pt}\\[0.7em]
{\large\bfseries End Matter}\\[0.3em]
\rule{0.42\textwidth}{0.4pt}
\end{center}
\vspace{0.6em}

\section{Finite-size extrapolation of the moments}
\label{app:extrap}

\onecolumngrid
\begin{figure}[h!]
\centering
\includegraphics[width=0.47\textwidth]{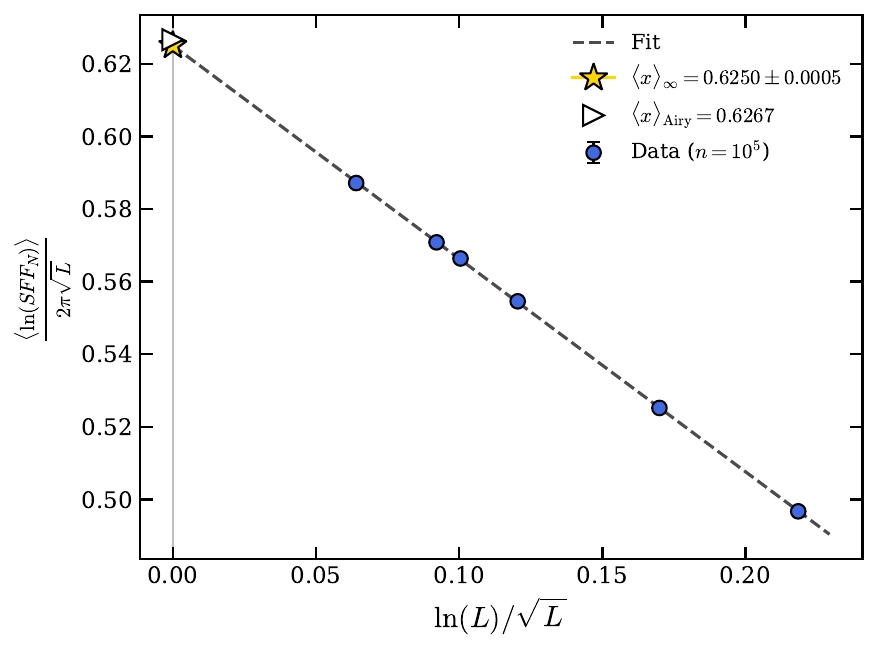}\hfill
\includegraphics[width=0.47\textwidth]{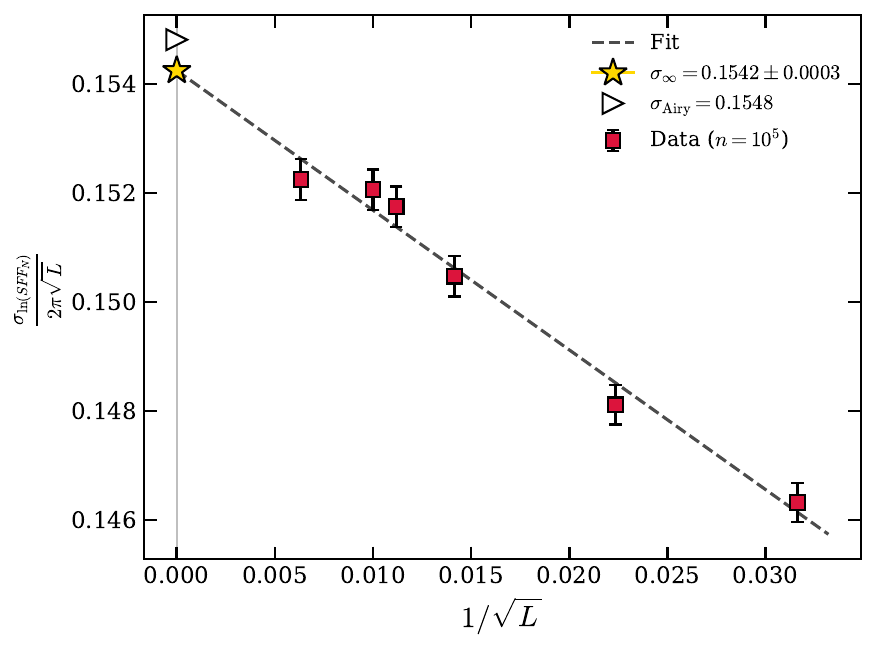}
\caption{Infinite-size extrapolation of the first two moments of $x=\ln \SFF_N /(2\pi\sqrt{L})$ for the Anderson model at $\rho=1/2$, $W=1$, from $10^5$ times sampled uniformly in $t\in[10^3,10^8]$. \emph{(a)} Mean, against $\ln L/\sqrt L$: $\langle x\rangle_\infty = 0.6250 \pm 0.0005$, to be compared with $\sqrt{\pi/8}=0.62666$ (triangle); error bars are invisible.  
\emph{(b)} Standard deviation, against $1/\sqrt L$: $\sigma_\infty = 0.1542 \pm 0.0003$, against $\sigma_{\rm Airy} = 0.15482$. No parameter is adjusted in the comparison.}
\label{fig:scaling}
\end{figure}
\twocolumngrid

We consider the rescaled variable $x=\ln\SFF_N/(2\pi\sqrt{L})$, which by Eq.~(\ref{eq:finalresult}) is Airy distributed as $L\to\infty$. 

In order to establish a finite size analysis we take into account the relative correction of order $\ln L/\sqrt{L}$, so that the mean
$\langle x\rangle$ and the standard deviation $\sigma_x$ are expected to behave as
\beq
\langle x \rangle = \sqrt{\tfrac{\pi}{8}} + a\,\frac{\ln L}{\sqrt{L}} + O\!\Big(\tfrac{1}{\sqrt{L}}\Big),
\quad
\sigma_x = \sigma_{\rm Airy} + O\!\Big(\tfrac{1}{\sqrt{L}}\Big).
\label{eq:moments}
\eeq
with $\sigma_{\rm Airy} \simeq 0.1548$.
Figure~\ref{fig:scaling} shows the first two moments of $x$ plotted against the variables prescribed by the structure of the corrections in Eq.~\eqref{eq:moments}: $\ln L/\sqrt L$ for the mean and $1/\sqrt L$ for the standard deviation, both vanishing in the thermodynamic limit. The data align on straight lines  
which extrapolate to $\langle x\rangle_\infty = 0.6250 \pm 0.0005$ and $\sigma_\infty = 0.1542\pm0.0003$, close to the predictions from the Airy distribution.


\clearpage
\onecolumngrid
\begin{center}
{\large\bfseries Supplemental Material for\\[0.3em]
``Airy distribution of the quenched spectral form factor in non-interacting fermions''}\\[0.6em]
Oscar Martin, Satya N. Majumdar, and Laura Foini
\end{center}
\vspace{0.6em}

In this Supplemental Material we give the details of the steps summarized in the main text: the replacement of the time average by an average over i.i.d.\ phases (Sec.~\ref{sm:weyl}), the evaluation of the Cauchy integral \eqref{eq:cauchy} (Sec.~\ref{sm:laplace}), and the covariance of the landscape and of its increments (Sec.~\ref{sm:cov}). We then recall why rationally independent many-body levels lead to a reflected Gumbel law for $\ln\SFF$ (Sec.~\ref{sm:ratind}), and show that the many-body levels of non-interacting fermions are rationally dependent, so that this argument does not apply to them (Sec.~\ref{sm:free}).

\setcounter{equation}{0}
\setcounter{figure}{0}
\setcounter{section}{0}
\setcounter{secnumdepth}{3}
\renewcommand{\theequation}{S\arabic{equation}}
\renewcommand{\thefigure}{S\arabic{figure}}
\renewcommand{\thesection}{S\Roman{section}}
\renewcommand{\thesubsection}{\thesection.\Alph{subsection}}

\section{Long-time average and equidistribution}
\label{sm:weyl}

At fixed disorder, $\SFF_N(t)$ depends on time only through the vector of phases $\boldsymbol{\phi}(t)=(\phi_1(t),\dots,\phi_L(t))$, $\phi_j(t)=-\varepsilon_j t \bmod 2\pi$, which moves linearly on the torus $\mathbb{T}^L=[0,2\pi)^L$. 
The continuous-time, multidimensional version of Weyl's criterion \cite[Ch.~1, Thm.~9.1 and 9.9; see also Sec.~6]{KuipersNiederreiter} states that the trajectory is equidistributed:
\beq
\lim_{T\to\infty}\frac{1}{T}\int_0^T F\big(\boldsymbol{\phi}(t)\big)\,dt=\frac{1}{(2\pi)^L}\int_{\mathbb{T}^L} F(\boldsymbol{\phi})\,d\boldsymbol{\phi},
\label{eq:SMweyl}
\eeq
if and only if it holds for all Fourier modes $F=e^{i\boldsymbol{m}\cdot\boldsymbol{\phi}}$, $\boldsymbol{m}\in\mathbb{Z}^L\setminus\{\boldsymbol 0\}$. For such a mode
\beq
\frac{1}{T}\int_0^T e^{-i(\boldsymbol{m}\cdot\boldsymbol{\varepsilon})t}\,dt
=\frac{1-e^{-i(\boldsymbol{m}\cdot\boldsymbol{\varepsilon})T}}{i(\boldsymbol{m}\cdot\boldsymbol{\varepsilon})T}\xrightarrow[T\to\infty]{}0,
\eeq
provided $\boldsymbol{m}\cdot\boldsymbol{\varepsilon}=\sum_j m_j\varepsilon_j\neq0$. The condition required is therefore that the $\varepsilon_j$ be \emph{rationally independent}: no nontrivial integer combination of them vanishes. Under this condition, Eq.~\eqref{eq:SMweyl} states that sampling $t$ uniformly at long times is equivalent to drawing $\boldsymbol\phi$ from the uniform measure on $\mathbb{T}^L$, $p(\phi_1,\dots,\phi_L)=\prod_j (2\pi)^{-1}$, i.e.\ to the i.i.d.\ uniform phases of Eq.~\eqref{eq:iidphases}. The convergence in Eq.~\eqref{eq:SMweyl} applies to the distribution of $\ln\SFF_N$, by taking for $F$ the indicator function of the event $\ln\SFF_N\le y$.

\section{Evaluation of the Cauchy integral}
\label{sm:laplace}

\subsection{Modulus and phase of the integrand}

With $a_j=e^{i\phi_j}$, Eq.~\eqref{eq:cauchy} reads
\beq
Z_N=\frac{1}{2\pi}\int_0^{2\pi} e^{F(\theta)}\,d\theta,\qquad
F(\theta)=\sum_{j=1}^L\ln\!\big(1+a_je^{i\theta}\big)-iN\theta = g(\theta)+i\Psi(\theta).
\label{eq:SMF}
\eeq
For a single factor, with $x=\theta+\phi_j$,
\beq
1+e^{ix}=2\cos\!\Big(\frac x2\Big)\,e^{ix/2},
\qquad\text{so that}\qquad
\ln|1+e^{ix}|=\ln\Big|2\cos\frac x2\Big|,\qquad \arg(1+e^{ix})=\frac x2 \quad (|x|<\pi).
\eeq
Differentiating,
\beq
\frac{d}{d\theta}\ln\!\big(1+a_je^{i\theta}\big)=\frac{i\,a_je^{i\theta}}{1+a_je^{i\theta}}
=\frac{i}{2}-\frac12\tan\!\Big(\frac{\theta+\phi_j}{2}\Big).
\label{eq:SMderiv}
\eeq
The imaginary part is the same constant $1/2$ for every $j$, which gives
\beq
\Psi'(\theta)=\frac L2-N=L\Big(\frac12-\rho\Big)\equiv\kappa
\label{eq:SMkappa}
\eeq
everywhere except at the zeros $\theta_j=\pi-\phi_j$ of $P(e^{i\theta})$, where the factor $1+a_je^{i\theta}\simeq -i(\theta-\theta_j)$ changes sign: as $\theta$ crosses $\theta_j$ the argument of that factor jumps from $+\pi/2$ to $-\pi/2$, so that
\beq
\Psi(\theta_j^-)-\Psi(\theta_j^+)=\pi .
\eeq
All other factors are smooth at $\theta_j$. Ordering the zeros along the circle, $\theta_{(1)}<\theta_{(2)}<\dots<\theta_{(L)}$, and denoting by $A_k=(\theta_{(k)},\theta_{(k+1)})$ the $L$ arcs between consecutive zeros (with $\theta_{(L+1)}=\theta_{(1)}+2\pi$), the phase is piecewise linear,
\beq
\Psi(\theta)=\kappa\,\theta-k\pi+\Psi_0,\qquad \theta\in A_k,
\eeq
with a constant $\Psi_0$. The integral therefore decomposes exactly into a sum over arcs with alternating signs:
\beq
Z_N=\frac{e^{i\Psi_0}}{2\pi}\sum_{k=1}^{L}(-1)^k I_k,\qquad
I_k=\int_{A_k}e^{g(\theta)}\,e^{i\kappa\theta}\,d\theta .
\label{eq:SMarcs}
\eeq
At half filling, $\kappa=0$ and each $I_k$ is real and positive.

\subsection{Concavity of the landscape on each arc}

From Eq.~\eqref{eq:SMderiv},
\beq
g'(\theta)=-\frac12\sum_{j=1}^L\tan\!\Big(\frac{\theta+\phi_j}{2}\Big),\qquad
g''(\theta)=-\frac14\sum_{j=1}^L\frac{1}{\cos^2\!\big(\frac{\theta+\phi_j}{2}\big)}
=-\frac14\sum_{j=1}^L\frac{1}{\sin^2\!\big(\frac{\theta-\theta_j}{2}\big)} .
\label{eq:SMgpp}
\eeq
Hence $g''<0$ everywhere: $g$ is strictly concave on each arc $A_k$, and since $g\to-\infty$ at both ends of the arc, it has there a unique critical point $\theta_k^*$, which is a maximum, with height $g_k^*=g(\theta_k^*)$. The global maximum of the landscape is $\max_\theta g=\max_k g_k^*$. The fine structure of $g$ is thus a sequence of $L$ concave bumps, separated by the logarithmic downward spikes, whose tops follow the Brownian envelope.

The curvature at the top of a bump is set by the nearest zeros. Since the $\theta_j$ are i.i.d.\ uniform on the circle, the arc lengths $\lambda_k=|A_k|$ are, for large $L$, of the form $\lambda_k=(2\pi/L)\,e_k$ with $e_k$ exponential random variables of unit mean. Each term in Eq.~\eqref{eq:SMgpp} behaves as $(\theta-\theta_j)^{-2}$ at short distance, and the zeros at distance $\sim 2\pi m/L$ contribute $\sim L^2/(4\pi^2m^2)$, a convergent sum. Therefore
\beq
|g''(\theta_k^*)|=c_k\,L^2,
\label{eq:SMck}
\eeq
where $c_k$ is a random variable of order one, depending on the local configuration of zeros only; in particular $c_k\ge 4/(L\lambda_k)^2$, since the distance from $\theta_k^*$ to the nearest end of the arc is at most $\lambda_k/2$.

The scaling \eqref{eq:SMck} is consistent with the Brownian character of the landscape (Sec.~\ref{sm:cov}). A bump extends over an arc of typical length $\delta\simeq2\pi/L$, across which $g$ varies by an amount of order the standard deviation of its increments, $\langle[g(\theta+\delta)-g(\theta)]^2\rangle^{1/2}\simeq(\pi L\delta/2)^{1/2}=\pi$, i.e.\ of order one. The second-moment formula is exact for i.i.d.\ phases at any $L$; only the Gaussian character of the increments fails on this scale. Matching this $O(1)$ variation to the quadratic approximation, $\tfrac12|g''(\theta_k^*)|\,\delta^2=O(1)$, gives $|g''(\theta_k^*)|=O(1/\delta^2)=O(L^2)$, as in Eq.~\eqref{eq:SMck}.

\subsection{Laplace evaluation of a single arc}

Expanding around the top of the bump, $g(\theta)\simeq g_k^*-\tfrac12|g''(\theta_k^*)|(\theta-\theta_k^*)^2$, the Gaussian integral gives
\beq
I_k\simeq e^{g_k^*+i\kappa\theta_k^*}\sqrt{\frac{2\pi}{|g''(\theta_k^*)|}}\;
\exp\!\left(-\frac{\kappa^2}{2|g''(\theta_k^*)|}\right)
=\frac{e^{g_k^*+i\kappa\theta_k^*}}{L}\sqrt{\frac{2\pi}{c_k}}\;
\exp\!\left(-\frac{(\frac12-\rho)^2}{2c_k}\right).
\label{eq:SMlaplace}
\eeq
Two features of Eq.~\eqref{eq:SMlaplace} are essential.
\begin{itemize}
\item[(a)] \emph{The width of the bump is $O(1/L)$}, which produces the factor $1/L$ and hence the logarithmic correction $-\ln L$ to $\ln|Z_N|$.
\item[(b)] \emph{The phase cannot cancel the bump.} Across its width the phase $\kappa\theta$ turns by $\kappa/(L\sqrt{c_k})=(\tfrac12-\rho)/\sqrt{c_k}=O(1)$; correspondingly, the oscillating factor reduces $I_k$ only by the $L$-independent factor $\exp[-(\tfrac12-\rho)^2/2c_k]$. This is the statement, made in the main text, that the dense set of zeros on the unit circle makes each peak too narrow to be destroyed by the phase $e^{-iN\theta}$, even though $N\propto L$.
\end{itemize}
The Gaussian approximation is not parametrically controlled, since the width of the Gaussian and the length of the arc are both $O(1/L)$. The integral can nevertheless be controlled by zooming on the global maximum of the landscape. Let $\theta^*$ be the position of the global maximum, $g^*=g(\theta^*)=\max_\theta g=\max_k g_k^*$, i.e.\ the top of the highest bump. On scales $\delta\gg1/L$, $g$ follows its Brownian envelope, and a Brownian path seen from its maximum lies below it by an amount of the order of the standard deviation of its increments: with the increment variance $\pi L\delta/2$ of Sec.~\ref{sm:cov}, $g^*-g(\theta^*\pm\delta)\sim\sqrt{L\delta}$, a square-root rather than quadratic decrease. Hence $e^{g}$ drops by $O(1)$ over $\delta\sim1/L$. This matches the width of a single concave bump, Eq.~\eqref{eq:SMck}, and is also the spacing of the zeros and the scale over which the phase $\kappa\theta$ turns by $O(1)$: there is a single microscopic scale. Setting $\theta=\theta^*+u/L$ and $h_L(u)=g(\theta^*+u/L)-g^*$, we get 
\beq
Z_N=\frac{e^{g^*}}{L}\,R_L,\qquad
R_L=\frac{e^{i\Psi_0'}}{2\pi}\int du\;e^{h_L(u)}\,e^{i(\frac12-\rho)u}\,(-1)^{k(u)},
\label{eq:SMmicro}
\eeq
where $k(u)$ counts the zeros crossed and $\Psi_0'$ is a constant phase. Here $h_L\le0$ is of order one: it is quadratic within the central bump, decays typically as $-\sqrt{\pi|u|/2}$ for $|u|\gg1$, and carries logarithmic spikes at the zeros, which form a point process of unit density in $u/2\pi$. No large parameter is left in $R_L$. The factor $1/L$, and hence the $-\ln L$ correction, is simply the Jacobian $d\theta=du/L$.

The structure of $R_L$ is transparent arc by arc. The central arc contributes the term given explicitly by Eq.~\eqref{eq:SMlaplace}, of order one. Neighbouring bumps, $2\pi/L$ apart, differ in height by $O(1)$ and contribute comparably, but the bump $m$ arcs away lies typically $\sim\pi\sqrt m$ below the maximum, so these contributions decay as $e^{-\pi\sqrt m}$ and only a number of order one of them matter. Bumps belonging to other excursions of the envelope are suppressed by factors $e^{-O(\sqrt L)}$. We therefore expect $R_L$ to have an $L$-independent limiting distribution, set by the local structure of the landscape near its maximum. The alternating signs $(-1)^{k(u)}$ may occasionally produce near cancellations, making $\ln|R_L|$ large and negative. This is the counterpart, at this order, of the Fisher zeros responsible for the left tail of the Gumbel law, and it only affects the $O(1)$ term. Taking the logarithm,
\beq
\ln|Z_N|=\max_\theta g(\theta)-\ln L+\ln|R_L| ,
\label{eq:SMfinal}
\eeq
which is Eq.~\eqref{eq:lnZmax} of the main text, with an explicit $-\ln L$ and an $O(1)$ random remainder.

\section{Covariance of the landscape and of its increments}
\label{sm:cov}

\subsection{Two-point function}

Since the phases are i.i.d.\ and each term of $g$ has zero mean, only the diagonal terms survive in $C(\tau)=\langle g(\theta)g(\theta+\tau)\rangle$:
\beq
C(\tau)=L\int_0^{2\pi}\frac{d\phi}{2\pi}\,
\ln\big|1+e^{i(\theta+\phi)}\big|\,\ln\big|1+e^{i(\theta+\tau+\phi)}\big| .
\eeq
Using the Fourier series $\ln|1+e^{ix}|=\sum_{n\ge1}\frac{(-1)^{n-1}}{n}\cos(nx)$ and $\int_0^{2\pi}\frac{d\phi}{2\pi}\cos(n(\phi+\theta))\cos(n'(\phi+\theta+\tau))=\tfrac12\delta_{nn'}\cos(n\tau)$,
\beq
C(\tau)=\frac L2\sum_{n\ge1}\frac{\cos(n\tau)}{n^2}
=L\left(\frac{\pi^2}{12}-\frac{\pi}{4}|\tau|+\frac{\tau^2}{8}\right)
=\frac{\pi^2L}{2}\,B_2\!\Big(\frac{\tau}{2\pi}\Big),\qquad |\tau|\le2\pi,
\label{eq:SMcov}
\eeq
with $B_2(s)=s^2-s+\tfrac16$. Note that $\int_0^{2\pi}C(\tau)\,d\tau=0$, consistently with the zero-area sum rule.

\subsection{Increments}

For a single increment $\Delta=g(\theta+\delta)-g(\theta)$,
\beq
\langle\Delta^2\rangle=2\big[C(0)-C(\delta)\big]=L\left(\frac{\pi}{2}|\delta|-\frac{\delta^2}{4}\right)\simeq \frac{\pi L}{2}|\delta|,\qquad |\delta|\ll1,
\eeq
the variance of a Brownian motion with diffusion constant $\pi L/4$. For two disjoint increments $\Delta_1=g(\theta_2)-g(\theta_1)$ and $\Delta_2=g(\theta_4)-g(\theta_3)$, with $\theta_1<\theta_2\le\theta_3<\theta_4$ and $\theta_4-\theta_1\le2\pi$,
\beq
\begin{split}
\langle\Delta_1\Delta_2\rangle
&=C(\theta_4-\theta_2)-C(\theta_3-\theta_2)-C(\theta_4-\theta_1)+C(\theta_3-\theta_1).
\end{split}
\eeq
The constant and linear parts of $C$ cancel in this combination, and the quadratic part gives
\beq
\langle\Delta_1\Delta_2\rangle=-\frac{L}{4}\,(\theta_2-\theta_1)(\theta_4-\theta_3).
\label{eq:SMincr}
\eeq
The increments are therefore not independent: disjoint increments are weakly and uniformly anticorrelated. For infinitesimal increments, normalized by the diffusive scale, $\langle\Delta_1\Delta_2\rangle/\big(\tfrac{\pi L}{2}\big)=-d\theta_1\,d\theta_2/(2\pi)$, a correlation of order $d\theta_1 d\theta_2$, much smaller than the variances $\propto d\theta$, which leaves the local Brownian character untouched. This uniform anticorrelation is precisely the one of a periodic Brownian path with its spatial average subtracted: for $\mathcal{B}_0(u)=B(u)-\int_0^1B$ one finds in the same way $\langle d\mathcal B_0(u_1)\,d\mathcal B_0(u_2)\rangle=-du_1du_2$ for $u_1\ne u_2$, which coincides with Eq.~\eqref{eq:SMincr} under $g(2\pi u)=\pi\sqrt L\,\mathcal B_0(u)$. It enforces the global constraint that the integral of the path vanishes, which for $g$ holds exactly.

\section{Rationally independent many-body levels: reflected Gumbel law}
\label{sm:ratind}

Consider $\D$ many-body levels $E_1,\dots,E_{\D}$ that are rationally independent,
\beq
\sum_{n=1}^{\D} m_n E_n \neq 0 \qquad \text{for every } (m_1,\dots,m_{\D})\in\mathbb{Z}^{\D}\setminus\{\boldsymbol 0\},
\eeq
and the partition function at $\beta=0$,
\beq
Z(it)=\sum_{n=1}^{\D} e^{-iE_n t}=X_R+iX_I,\qquad
X_R=\sum_{n=1}^{\D}\cos(E_n t),\quad X_I=-\sum_{n=1}^{\D}\sin(E_n t).
\eeq
Applying the Weyl criterion of Sec.~\ref{sm:weyl} to the $\D$ levels (instead of the $L$ single-particle energies), sampling $t$ uniformly at long times is equivalent to replacing $E_nt \bmod 2\pi$ by independent phases $\varphi_n$, uniform on $[0,2\pi)$. Since $\langle\cos^2\varphi_n\rangle=\langle\sin^2\varphi_n\rangle=1/2$ and $\langle\cos\varphi_n\sin\varphi_n\rangle=0$,
\beq
\langle X_R\rangle=\langle X_I\rangle=0,\qquad
\langle X_R^2\rangle=\langle X_I^2\rangle=\frac{\D}{2},\qquad
\langle X_RX_I\rangle=0 .
\eeq
By the multivariate central limit theorem, for large $\D$ the pair $(X_R,X_I)$ converges to two independent Gaussian variables:
\beq
P(x_R,x_I)\to\frac{1}{\pi\D}\,e^{-(x_R^2+x_I^2)/\D}.
\label{eq:SMgauss}
\eeq
The SFF, $\SFF=X_R^2+X_I^2$, is then exponentially distributed (pass to polar coordinates in Eq.~\eqref{eq:SMgauss}):
\beq
P_{\SFF}(s)=\frac{1}{\D}\,e^{-s/\D},\qquad s\ge0 .
\label{eq:SMexp}
\eeq

\paragraph{The SFF is not self-averaging.}
From Eq.~\eqref{eq:SMexp},
\beq
\langle\SFF\rangle=\int_0^\infty s\,P_{\SFF}(s)\,ds=\D,\qquad
\langle\SFF^2\rangle=2\D^2,
\eeq
so that $\sqrt{\langle\SFF^2\rangle-\langle\SFF\rangle^2}=\D=\langle\SFF\rangle$: the fluctuations are as large as the mean, whatever $\D$. 

\paragraph{Its logarithm is self-averaging.}
Let $\zeta=\ln\SFF$. Changing variables to $y=s/\D$,
\beq
\langle\zeta\rangle=\int_0^\infty \ln s\,\frac{e^{-s/\D}}{\D}\,ds
=\ln\D+\int_0^\infty \ln y\,e^{-y}\,dy=\ln\D-\gamma,
\eeq
\beq
\langle\zeta^2\rangle-\langle\zeta\rangle^2
=\int_0^\infty(\ln y)^2e^{-y}\,dy-\Big(\int_0^\infty\ln y\,e^{-y}\,dy\Big)^2
=\frac{\pi^2}{6},
\eeq
with $\gamma$ the Euler-Mascheroni constant. The standard deviation of $\zeta$ is $O(1)$, while its mean grows as $\ln\D\propto L$: the relative fluctuations vanish, $\pi/(\sqrt6\ln\D)\to0$, and $\zeta$ is self-averaging.

\paragraph{Reflected Gumbel law.}
For the logarithm of the scaled variable,
\beq
Y=\ln\frac{\SFF}{\D},\qquad
P(Y>y)=P\big(\SFF>\D e^{y}\big)=\exp\!\big(-e^{y}\big),
\eeq
so that the density of $Y$ is the reflected Gumbel law,
\beq
P_Y(y)=e^{\,y-e^{y}},
\label{eq:SMgumbel}
\eeq
with mean $-\gamma$ and variance $\pi^2/6$. This is the distribution shown for the XXZ chain in Fig.~\ref{fig:xxzgumbel} of the main text.

\section{Rational dependence of free-fermion many-body levels}
\label{sm:free}

For non-interacting fermions, the many-body levels are rationally dependent even when the single-particle energies are rationally independent, so the argument of Sec.~\ref{sm:ratind} does not apply.

\paragraph{A minimal example.}
Take $L=4$ rationally independent single-particle energies $\varepsilon_1,\dots,\varepsilon_4$ and $N=2$ fermions. The $\D=\binom42=6$ many-body levels are $E_{ij}=\varepsilon_i+\varepsilon_j$, $i<j$, and they satisfy
\beq
E_{12}+E_{34}=E_{13}+E_{24}=E_{14}+E_{23}=\sum_{j=1}^{4}\varepsilon_j ,
\eeq
i.e.\ two independent integer relations among the six levels. The Weyl criterion therefore fails for the many-body phases $e^{-iE_{ij}t}$, although it holds for the single-particle ones.

\paragraph{General case.}
Every many-body level $E_{\boldsymbol n}=\sum_j n_j\varepsilon_j$, $n_j\in\{0,1\}$, is an integer combination of the same $L$ single-particle energies. Hence at most $L$ of the $\D$ levels can be rationally independent, and the others obey integer relations among themselves: at least $\D-L$ independent ones. Since $\D\simeq e^{Ls(\rho)}$, almost all of them are rationally dependent. Correspondingly, the many-body phases $e^{-iE_{\boldsymbol n}t}\to\prod_j a_j^{n_j}$ are products of the same $L$ independent phases $a_j$ of Eq.~\eqref{eq:iidphases}, and are strongly correlated.

\end{document}